\documentclass{wscpaperproc}
\usepackage{latexsym}
\usepackage{graphicx}
\usepackage{mathptmx}
\usepackage[T1]{fontenc}

\usepackage{amsmath}
\usepackage{amsfonts}
\usepackage{amssymb}
\usepackage{amsbsy}
\usepackage{amsthm}
\usepackage[pdftex,colorlinks=true,urlcolor=blue,citecolor=black,anchorcolor=black,linkcolor=black]{hyperref}

\usepackage[pdftex,colorlinks=true,urlcolor=blue,citecolor=black,anchorcolor=black,linkcolor=black]{hyperref}

\usepackage{booktabs}
\usepackage{subcaption}
\usepackage{cleveref}

\newtheoremstyle{wsc}
{3pt}
{3pt}
{}
{}
{\bf}
{}
{.5em}
{}

\theoremstyle{wsc}

\begin{document}

%
%

\pagestyle{fancyplain}

\thispagestyle{plain}
\firstPageHead{}

\chead{\fancyplain{}{\itshape Giabbanelli}}

\rhead{}
\cfoot{}
\renewcommand{\headrulewidth}{0pt} 

\makeatletter
\let\@internalcite\cite
\def\cite{\def\@citeseppen{-1000}%
    \def\@cite##1##2{(##1\if@tempswa , ##2\fi)}%
    \def\citeauthoryear##1##2##3{##1 ##3}\@internalcite}
\def\citeNP{\def\@citeseppen{-1000}%
    \def\@cite##1##2{##1\if@tempswa , ##2\fi}%
    \def\citeauthoryear##1##2##3{##1 ##3}\@internalcite}
\def\citeN{\def\@citeseppen{-1000}%
    \def\@cite##1##2{##1\if@tempswa, ##2)\else{}\fi}%
    \def\citeauthoryear##1##2##3{##1 (##3)}\@citedata}
\def\citeA{\def\@citeseppen{-1000}%
    \def\@cite##1##2{(##1\if@tempswa , ##2\fi)}%
    \def\citeauthoryear##1##2##3{##1}\@internalcite}
\def\citeANP{\def\@citeseppen{-1000}%
    \def\@cite##1##2{##1\if@tempswa , ##2\fi}%
    \def\citeauthoryear##1##2##3{##1}\@internalcite}
\def\shortcite{\def\@citeseppen{-1000}%
    \def\@cite##1##2{(##1\if@tempswa , ##2\fi)}%
    \def\citeauthoryear##1##2##3{##2 ##3}\@internalcite}
\def\shortciteNP{\def\@citeseppen{-1000}%
    \def\@cite##1##2{##1\if@tempswa , ##2\fi}%
    \def\citeauthoryear##1##2##3{##2 ##3}\@internalcite}
\def\shortciteN{\def\@citeseppen{-1000}%
    \def\@cite##1##2{##1\if@tempswa, ##2\else{}\fi}%
    \def\citeauthoryear##1##2##3{##2 (##3)}\@citedata}
\def\shortciteA{\def\@citeseppen{-1000}%
    \def\@cite##1##2{(##1\if@tempswa , ##2\fi)}%
    \def\citeauthoryear##1##2##3{##2}\@internalcite}
\def\shortciteANP{\def\@citeseppen{-1000}%
    \def\@cite##1##2{##1\if@tempswa , ##2\fi}%
    \def\citeauthoryear##1##2##3{##2}\@internalcite}
\def\citeyear{\def\@citeseppen{-1000}%
    \def\@cite##1##2{(##1\if@tempswa , ##2\fi)}%
    \def\citeauthoryear##1##2##3{##3}\@citedata}
\def\citeyearNP{\def\@citeseppen{-1000}%
    \def\@cite##1##2{##1\if@tempswa , ##2\fi}%
    \def\citeauthoryear##1##2##3{##3}\@citedata}
%
%
%
\def\@citedata{%
    \@ifnextchar [{\@tempswatrue\@citedatax}%
                  {\@tempswafalse\@citedatax[]}%
}

\def\@citedatax[#1]#2{%
\if@filesw\immediate\write\@auxout{\string\citation{#2}}\fi%
  \def\@citea{}\@cite{\@for\@citeb:=#2\do%
    {\@citea\def\@citea{, }\@ifundefined
       {b@\@citeb}{{\bf ?}%
       \@warning{Citation `\@citeb' on page \thepage \space undefined}}%
{\csname b@\@citeb\endcsname}}}{#1}}%

%
\def\@citex[#1]#2{%
\if@filesw\immediate\write\@auxout{\string\citation{#2}}\fi%
  \def\@citea{}\@cite{\@for\@citeb:=#2\do%
    {\@citea\def\@citea{; }\@ifundefined
       {b@\@citeb}{{\bf ?}%
       \@warning{Citation `\@citeb' on page \thepage \space undefined}}%
{\csname b@\@citeb\endcsname}}}{#1}}%

%
\def\@biblabel#1{}
\makeatother



\newdimen\bibindent
\bibindent=0.0em
\def\thebibliography#1{\section*{\refname}\list
   {}{\settowidth\labelwidth{[#1]}
   \leftmargin\parindent
   \itemindent -\parindent
   \listparindent \itemindent
   \itemsep 0pt
   \parsep 0pt}
   \def\newblock{}
   \sloppy
   \sfcode`\.=1000\relax}


\setlength{\baselineskip}{12.7pt}

\title{GPT-BASED MODELS MEET SIMULATION: HOW TO EFFICIENTLY USE LARGE-SCALE PRE-TRAINED LANGUAGE MODELS ACROSS SIMULATION TASKS}

\author{Philippe J. Giabbanelli\\[12pt]
Department of Computer Science \& Software Engineering\\
Miami University\\
501 East High Street\\
Oxford, OH 45056, USA\\
}

\maketitle

\section*{ABSTRACT}
The disruptive technology provided by large-scale pre-trained language models (LLMs) such as ChatGPT or GPT-4 has received significant attention in several application domains, often with an emphasis on high-level opportunities and concerns. This paper is the first examination regarding the use of LLMs for scientific simulations. We focus on four modeling and simulation tasks, each time assessing the expected benefits and limitations of LLMs while providing practical guidance for modelers regarding the steps involved. The first task is devoted to explaining the structure of a conceptual model to promote the engagement of participants in the modeling process. The second task focuses on summarizing simulation outputs, so that model users can identify a preferred scenario. The third task seeks to broaden accessibility to simulation platforms by conveying the insights of simulation visualizations via text. Finally, the last task evokes the possibility of explaining simulation errors and providing guidance to resolve them.

\section{INTRODUCTION}
\label{sec:intro}
Natural Language Generation (NLG) has been in the limelight recently, following the release of ChatGPT and its wide potential application areas from writing (academic) papers to assignments and software code. Much in the same way as `Google' colloquially refers to using a search engine, ChatGPT has served as a proxy to discuss the opportunities and concerns raised by Large Language Models (LLMs)~\shortcite{van2023chatgpt,zhou2023chatgpt}. These large-scale pre-trained models are based on transformer architectures~\shortcite{tay2022efficient,wang2022pre} and include several versions of GPT (e.g., GPT4 with one trillion parameters) alongside Google's Pathways Language Model (PaLM, whose 540 billion parameters support the Bard chatbot), LLaMA from Meta (available at several sizes), or Megatron-Turing (530 billion parameters) created by Microsoft and NVIDIA~\shortcite{chowdhery2022palm,smith2022using}. Beyond the sensational headlines, there is a growing realization that these models are complex tools that require technical attention before being adequately deployed. As summarized by the editor-in-chief of \textit{Science}, ChatGPT provides ``endless entertainment'' but ultimately, like other machines, it serves as a tool ``for the people posing the hypotheses, designing the experiments, and making sense of the results''~\cite{thorp2023chatgpt}. For example, researchers illustrated that ChatGPT was not going to perform a literature review by itself, as two thirds of the scientific studies that it discussed did not exist~\cite{haman2023using}; this phenomenon known as \textit{hallucination} is one of the many errors or `unpredictable qualities' occurring in LLMs~\shortcite{ganguli2022predictability,borji2023categorical}. It is thus important to complement the nascent literature on high-level opportunities and concerns with an emphasis on \textit{practical tasks} and how they may be facilitated with LLMs under the right human intervention, which may include fine-tuning~\shortcite{ding2023parameter}, asking the right questions (i.e., prompt engineering as discussed in~\shortciteNP{white2023prompt}), and identifying where to correct the generated text.

In this paper, we are interested in shifting from using LLMs such as GPT as assistants in high-level science tasks (e.g., summarizing papers) to becoming central actors in specific tasks for Modeling \& Simulation (M\&S). This shift finds several parallels with the advent of Machine Learning and its impact on M\&S~\cite{giabbanelli2019solving,elbattah2019can}. Neither NLG nor machine learning are brand new, as their concepts and early systems were operational decades ago~\shortcite{gatt2018survey,dong2022survey}. However, their rise is based on the ability of new tools to operate at an unprecedented \textit{scale} while being easily \textit{accessible}. Plethora of online courses can equip practitioners with machine learning skills and a model can be quickly trained thanks to library such as {\ttfamily scikit-learn} or drag-and-drop software. Tools such as GPT have been in existence for several years already, as we presented a prototype using GPT-3 for simulation in 2022~\shortcite{shrestha2022automatically}. But the recent availability of products such as ChatGPT now makes these tools accessible, as there is no need for programming via an API. We can thus expect that NLG will potentially permeate most stages of the M\&S process, just as machine learning has become commonplace through a multitude of innovative hybrid simulations~\shortcite{muller2022towards,ghasemi2022demonstration,onggo2018symbiotic}. In this context, this paper contributes to \textit{preparing our research community for this shift by examining which M\&S tasks can benefit from NLG and how it would be achieved}. We note that such inventories of candidate tasks for NLG are now abundant in healthcare and medical education~\cite{sallam2023chatgpt}, business and marketing~\cite{rivas2023marketing}, or environmental science~\shortcite{zhu2023chatgpt}, but such guidance had yet to be issued for M\&S.

\begin{figure}[htb]
	\centering
	\includegraphics[width=\linewidth]{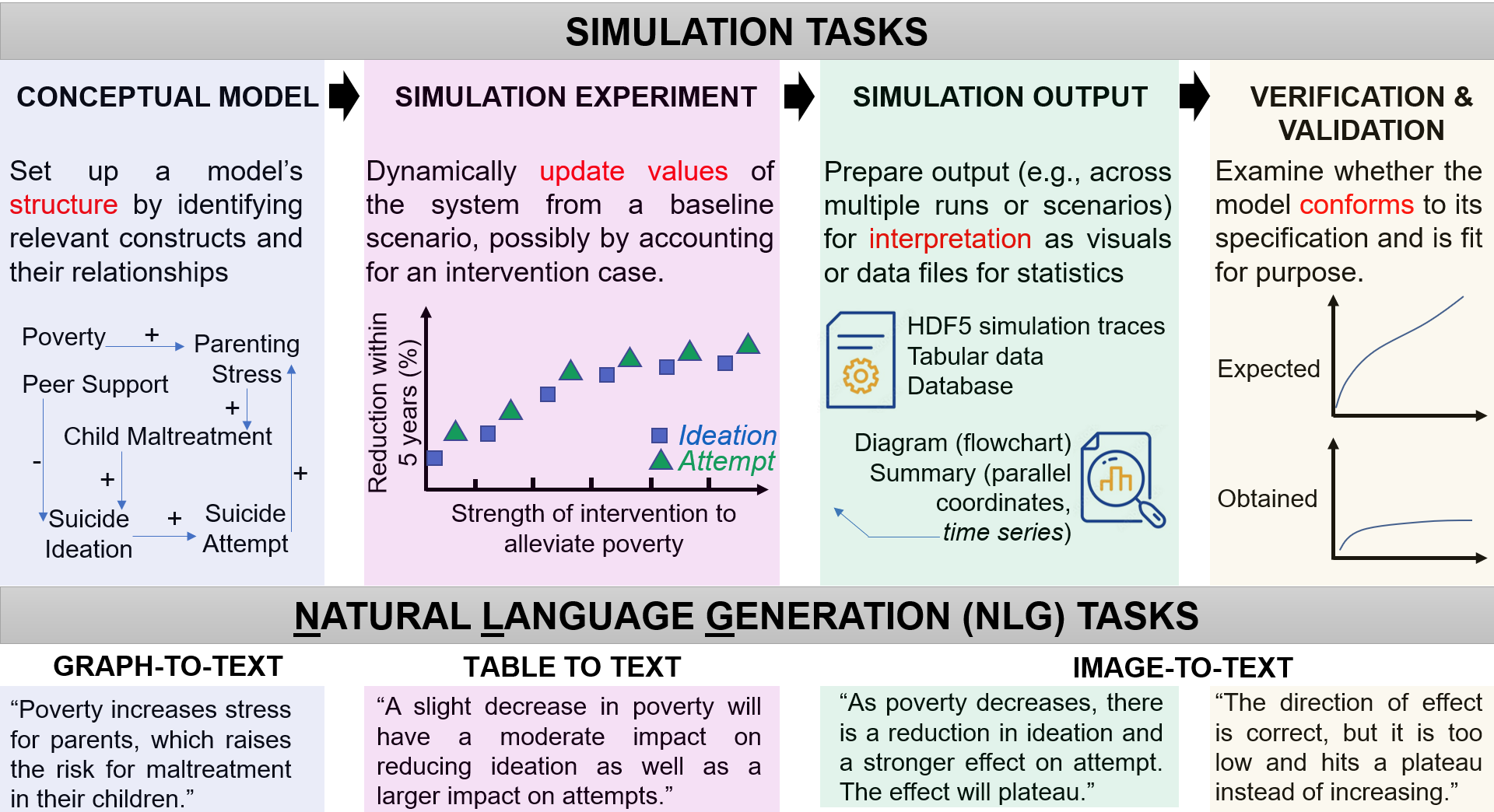}
  \caption{After creating a conceptual model, it is implemented and utilized to perform experiments whose outputs support decision-making activities. Outputs are also used to ensure the correctness of the model vis-à-vis its specification (i.e., verification) or the adequacy of the model with regards to its real-world counterpart (i.e., validation). In this paper, we map each of these steps to a task in NLG.}
	\label{fig:overview}
\end{figure}

This paper proceeds in the order of simulation tasks summarized in Figure~\ref{fig:overview}. This summary only illustrates the key steps of this paper, as we acknowledge that M\&S involves several other steps such as the transition from a conceptual model to a \textit{mathematical} specification and its \textit{implementation} as a computational model. As a popular phrase goes, ``if all you have is a hammer, everything looks like a nail''; each section thus begins by establishing the \textit{rationale} for using NLG. Then, we detail the \textit{methods} involved, while noting that their maturity decreases as we progress along simulation tasks.

\section{EXPLAINING THE STRUCTURE OF A SIMULATION MODEL}
\label{sec:structure}

\subsection{Rationale for Using Natural Language Generation}

Modelers work within interdisciplinary teams, where members have (potentially overlapping) roles such as model commissioners who set the purpose of the model or participants who inform its content~\shortcite{calder2018computational}. Empirical studies have repeatedly highlighted the importance of \textit{communication skills} in such teams. As discussed by~\citeN{ahrweiler2019co}, ``the first and most important [demand] is that the clients want to understand the model'', which means that the \textit{structure} of the model and the logic of its decision should be clear, rather than treated as blackbox that only allows to view and discuss outputs. Clearly conveying a model's structure is challenging, since team members may not be expert in modeling techniques hence they delving into code is not a viable solution. Although modeling languages (e.g., UML, SysML) can be familiar tools for model development, their steep learning curve for participants also presents an obstacle~\cite{padilla2019agent}. Our experiments confirmed that even a graph that only shows concepts and whether they are connected can be a significant learning curve for participants (senior executives), who struggled to provide confident and timely answers for basic questions about the model~\cite{giabbanelli2023human}. Natural Language Generation thus opens up the opportunity to explain a model in a format that is potentially accessible to all parties: textual narratives. 

\begin{figure}[htb]
	\centering
	\includegraphics[width=\linewidth]{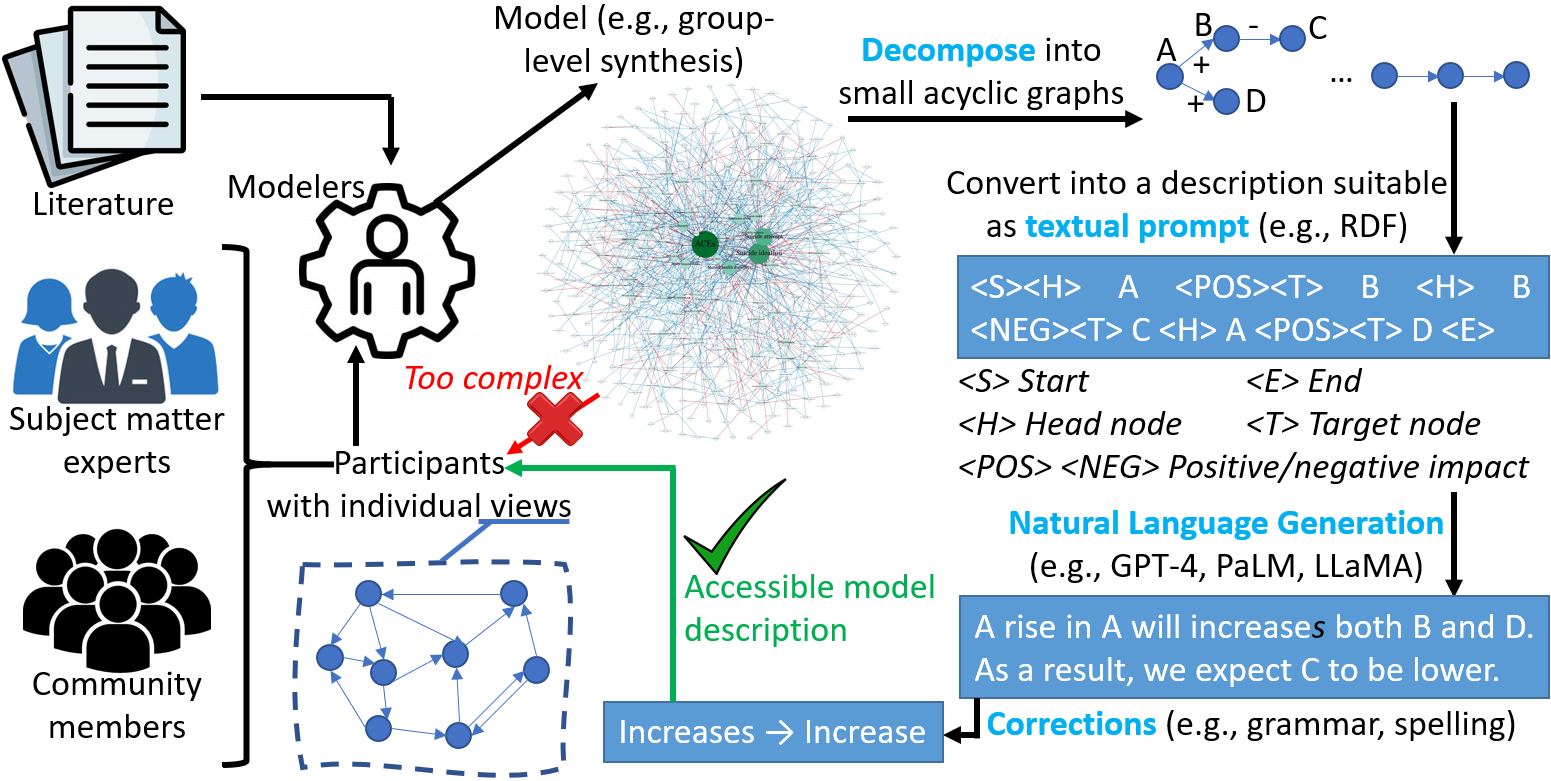}
  \caption{Modelers can work with the literature and/or participants to create a conceptual model. Each article or participant may provide a \textit{small} model on a facet of the problem, but the overall model may be larger and harder too explain due to the sheer number of constructs and relationships or complex dynamics (e.g., loops). NLG can explain the model in textual form and it requires several steps: decomposing the model into smaller inputs, converting them to textual prompts, and correcting the output (if necessary).}
	\label{fig:explaining}
\end{figure}

Explaining model as narratives has limitations. For example, modelers should still be involved to assist participants. As argued by~\shortciteN{gilbert2018computational}, ``it is impossible to capture in a report all the nuances of the model simplifications, data weaknesses etc. in a way that [participants] can use reliably''. In addition, a report automatically generated from a schema may not be able to cover all topics that ought to be communicated, such as the purpose of the model~\shortcite{grimm2020three}. However, we argue that an automatically generated report can at least convey the \textit{structure} of a model so that participants understand which variables are involved and how they interact. Providing these expectations can address the existing disconnect between the transparent and often informal process to \textit{elicit} information from participants, and the opacity of the resulting model (Figure~\ref{fig:explaining}). Since transparency and trust in a model often come together~\cite{falconi2017interdisciplinary}, we posit that a careful use of NLG to turn models into reports may ultimately increase the engagement of participants with the modeling process and their support of the decisions suggested by the simulations.

\subsection{Core Methodological Components}
\label{sec:explainSchema}

Although the relation between textual descriptions and conceptual models has been discussed elsewhere, it has primarily been from the viewpoint of extracting model elements from text~\cite{shuttleworth2022narratives}. This involves Natural Language \textit{Processing}, with a focus on identifying entities such as agents and their properties. In contrast, turning a model into text is a matter of Natural Language \textit{Generation} and it involves vastly different steps, particularly when operating via LLMs. Although LLMs such as GPT-4 have made headways with the use of images as inputs (Section~\ref{sec:visualizations}), modelers cannot generally expect to just `drop' a schema of their conceptual model and have it turned into a report. The schema first needs to be converted into a textual form.

Since schema (e.g., UML, causal maps) often depict concepts and their relations, the corresponding NLG task is known as \textit{graph-to-text}. Graphs can have cycles: for example, the conceptual model in Figure~\ref{fig:overview} has a cycle of parenting stress increasing the risk of suicide attempt in their children, thus causing more stress. In contrast, sentences must be linear. The conversion thus starts by turning the model's schema into linearized sentences, but it cannot simply be achieved by removing parts of the schema to break existing cycles (this is known as a `loss of structural information'). One strategy is to break the schema into parts that \textit{collectively} can recreate the full schema, thus resulting in some nodes being duplicated among the decomposed parts. These parts should be kept small (Figure~\ref{fig:explaining}) as experiments show that the quality of the generated text deteriorates with larger inputs~\shortcite{shrestha2022automatically}. Linearization is a problem in itself, as it cannot simply be achieved by removing parts of the schema to break existing cycles (this is known as a `loss of structural information'). One strategy is to break the schema into parts that \textit{collectively} can recreate the full schema, thus resulting in some nodes being duplicated among the decomposed parts. Alternatively, a new schema could be defined and introduce additional nodes to encapsulate the meaning of cycles~\cite{rodriguesgraph}. So far, linearization has primarily been studied for knowledge graphs rather than simulation schema, hence modelers would need to write custom linearization algorithms.

Once the graph is linearized, it needs to be expressed as a textual input. It does not suffice to just write out the graph as words, as the edges would be ambiguous. For example, the list $A, B, C, D$ could be interpreted as the edges $(A, B)$ and $(C, D)$, or $(A, B)$, $(B, C)$, $(C, D)$. An unambiguous representation thus uses flags to separate the origin node of an edge from the target. The Resource Description Framework (RDF) is widely used for this purpose~\shortcite{yang2020improving,zhang2023enhancing}. Depending on its sophistication, the LLM may need to be \textit{fine-tuned} by being presented with numerous RDF inputs and the expected sentence to generate. Once the (fine-tuned) LLM creates sentences, they may still need to be corrected. This can involve correcting typos (e.g., in earlier versions such as GPT-2), avoiding the redundancy that readers quickly associated with machine-generated text~\shortcite{liu2023argugpt}, or mitigating various forms of biases. The latter has received abundant attention, as authors have discussed the presence (and sometimes the inevitability) of bias~\cite{ferrara2023should} on sex, race, religion, or disability -- all of which are protected classes in US law on discrimination. However, studies on bias have not yet been conducted in the case of text generated from model schema, hence additional work is needed to assess the problem and whether some application fields are more at risk (e.g., models of social systems vs. physical systems).

Accomplishing the above steps results in generating \textit{sentences}, but it does not yet make a report. As an analogy, consider teaching: an instructor cannot deliver content material in random order, or teach a second year elective class in the same way as a graduate seminar. Generating a report also needs to account for the \textit{audience} and orchestrate sentences into meaningful \textit{paragraphs} with an appropriate flow. Satisfying either of these requirements is currently an open topic when applying NLG to explaining simulation models. Because modeling is conducted in an interdisciplinary setting, insufficiently accounting for ``the languages of the different research traditions can lead to misunderstanding and resentment''~\cite{smaldino2020translate}. We posit that forming paragraphs may be an easier problem (and hence a prime research target) because the readability of paragraphs can be automatically measured for a language in general (e.g., using the Flesch–Kincaid readability test), whereas the appropriateness of terms and style with regards to a scientific discipline does not have an algorithmic scoring method.



\section{HANDLING DYNAMICITY: COMPARING OUTCOMES FROM PREDICTIVE SIMULATIONS}
\label{sec:simulation}

\subsection{Rationale for Using Natural Language Generation}
\begin{figure}[htb]
	\centering
	\includegraphics[width=\linewidth]{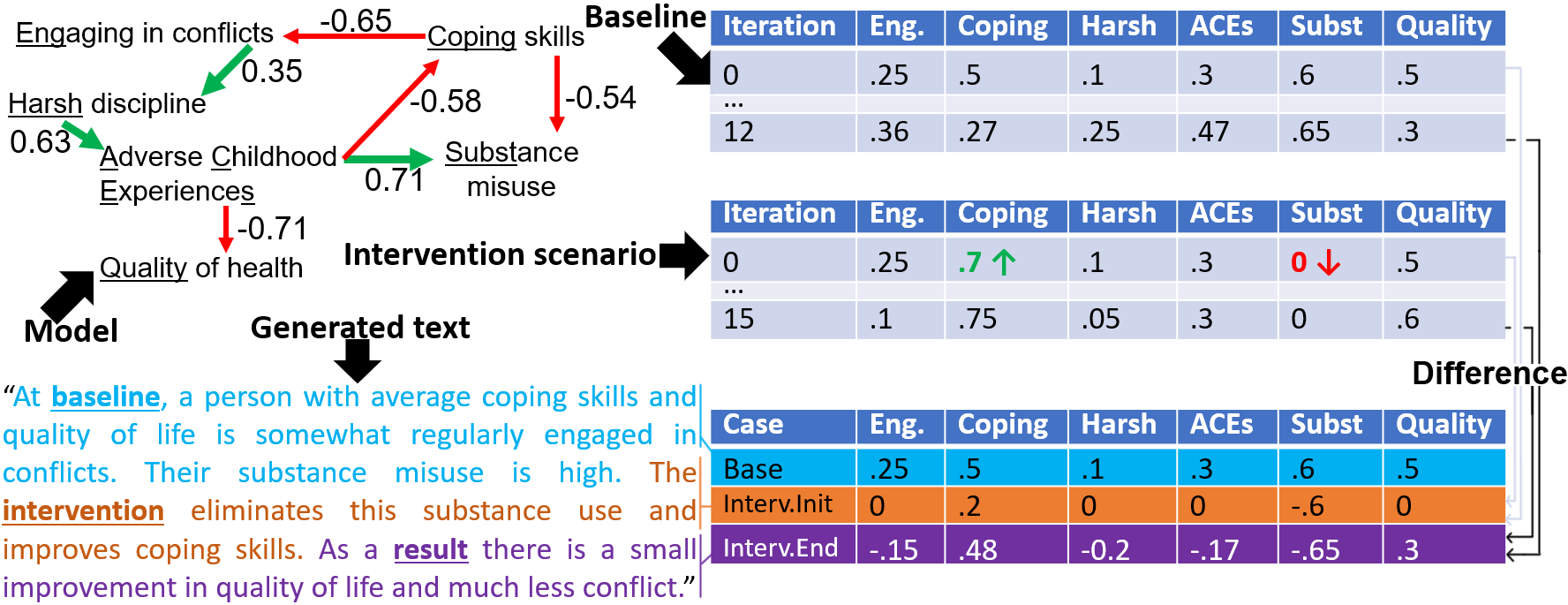}
  \caption{The illustrative model of Adverse Childhood Experiences (ACEs) is a part of a larger model on youth suicide, detailed in \protect\shortciteN{giabbanelli2022pathways}. The implications of the baseline or `status quo' scenario can be organized as a table, showing how the 6 constructs of the ACEs model change over discrete iterations until a final step. An intervention also consists of a table, whereby the initial values of some constructs change (due to the intervention) hence the final values may also change.}
	\label{fig:scenario}
\end{figure}

In the previous section, we discussed the benefits of explaining the \textit{structure} of a model by converting its schema to text. At a high-level, explaining the \textit{simulation outcomes} would yield similar benefits, such as greater transparency and engagement. To further estimate benefits and limitations, it is necessary to precisely define the task of `explaining outcomes'. A simulation model may be presented with a number of cases, also known as \textit{scenarios} or what-if questions. For instance, in a model for suicide prevention (Figure~\ref{fig:scenario}), such scenarios could include educating parents to avoid harsh discipline, improving coping mechanisms in children, or providing treatment for substance misuse. One of the cases may be marked as baseline, thus reflecting the current state of the world in the absence of hypothetical interventions. The goal is to inform model users about the difference in simulation outcomes across cases, such that they can \textit{choose} the best candidate interventions. By summarizing simulation outcomes across cases to focus on key differences, NLG can reduce the cognitive efforts involved in decision-making.

There are potential limitations when attempting to automatically convey the main differences across cases. First, a \textit{textual format may not always be most efficient}. For example, if a simulation has only one critical output, then users may prefer a bar chart visualization (Figure~\ref{fig:viz}) that shows the output of interest (y-axis) across simulation cases (x-axis). Since a bar chart relies on preattentive visual properties such as line length~\cite{wolfe2019preattentive} to bring attention to data points that stand out (e.g., lowest, highest), users would quickly notice which case yields the minimal/maximal outcome and hence would be preferred. A well-constructed visualization would thus be more effective than text. Second, a simulation may have multiple objectives that cannot all be optimized. For example, interventions for obesity could be characterized by indicators related to physical health (e.g., type-2 diabetes, musculoskeletal disorders, hypertension) as well as mental health (e.g., self-esteem and body image). Different users may give more importance to some of these indicators, and these preferences may be implicit. The incorporation of implicit preferences in multi-objective optimization~\shortcite{cruz2017incorporation} is a complex problem and has yet to be studied in the context of summaries generated by NLG. This section thus focuses on cases that have multiple outputs of interest and assumes that their importance is either equal or can be explicitly quantified.


\subsection{Core Methodological Components}

The input is a set of simulation cases, each containing the value of the model's constructs from the initial to the final iteration. Consider without loss of generality that the output seeks to summarize the baseline scenario, the design of each intervention and how it leads to changes compared to the baseline (Figure~\ref{fig:scenario}); the remainder of this section would be similar if there was no `baseline' case. As explained in section~\ref{sec:explainSchema}, a LLM does not directly go from the input to the desired output: modelers need to perform at least two additional steps. The first step is to gather the results into a single table that contains the characteristics of each intervention and the final value of each construct (Figure~\ref{fig:scenario}, bottom right); characteristics and final values can be expressed as a difference with respect to the baseline (if applicable). By transforming simulation outputs into one table, we frame the problem as a \textit{table-to-text task}, which has been well studied.

The second step is to transform the table into textual input for a LLM. Early methods follow a `pipeline paradigm' in which a \textit{table transformation module} applies a \textit{template} to turn a table into text~\shortcite{gong2020tablegpt}. Although this approach has the advantage of being conceptually simple, it requires users to design a template. Furthemore, the output is limited (hence repetitive) due to a reliance on set templates and rules. Newer approaches use end-to-end methods based on neural networks~\shortcite{yang2021table}, but they depend on a large training set and may not be readily applicable to the specific context of a simulation model. We refer the reader to~\shortciteN{guo2023few} for an overview of current options, and to Table 4 by~\citeN{sharma2022innovations} for a summary of methods based on the application dataset. While many prior works are concerned with tables that contain words, this is not directly applicable to simulation traces since they consist of \textit{numerical outputs}. We thus recommend methods specifically designed for tables with numerical content, such as~\shortciteN{suadaa2021towards}.

\begin{figure}[htb]
	\centering
	\includegraphics[width=\linewidth]{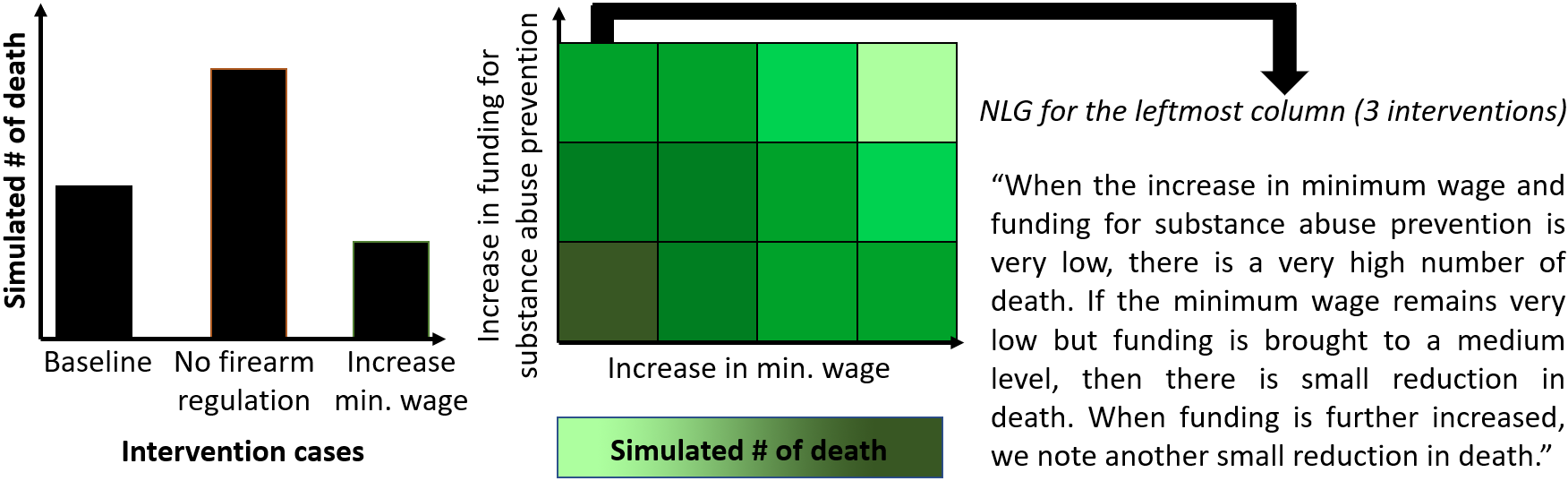}
  \caption{A simple bar chart (left) can quickly reveal the preferred scenario for a model user, as scanning the image to find the lowest/highest bar relies on preattentive visual features. Even when scenarios are not defined by intervention category (left) but rather than different levels of change in numerical parameters (center), a visualization can rely on other preattentive visual features such as hue to guide decision makers. In contrast, a complete text report can be overly verbose (right) and less effective for decision-making. There are thus situations in which visualizations can be used instead of, or in complement to, NLG.}
	\label{fig:viz}
\end{figure}

Accomplishing the above steps results in a \textit{complete} transformation of a table into text. If the table is short (e.g., few simulation cases and/or constructs), then model users may be able to read the generated text and identify the best simulation case. However, additional steps would be necessary to \textit{selectively} transform larger tables and avoid overly verbose reports (Figure~\ref{fig:viz}). As shown in Figure~\ref{fig:scenario}, it is not necessary to generate text regarding the initial state of every construct for every case: we can simply state which constructs had a different value than in the baseline case. We also do not need to specify the impact on every construct at the end of the simulation: applying a user-defined filter (e.g., ignore changes of less than 10\%, only include changes on three specific constructs) can trim the list. Beyond these simple means to keep the text short, we note that text summarization algorithms may offer additional solutions~\shortcite{gupta2019abstractive,raffel2020exploring}. These algorithms operate either by compiling the most important existing sentences (\textit{extractive} summarization e.g., Textrank, BERT-ext, Longformer-Ext) or by generating sentences (\textit{abstractive} summarization e.g., BART, T5), which tend to have higher readability and are more concise but may not exactly reflect the meaning of the original text~\shortcite{alomari2022deep}. However, new summarization algorithms would need to be developed since the existing ones are not readily applicable to simulation data, which consist entirely of numbers. Indeed, general purpose summarization algorithms either ignore numbers or treat sentences with numerical data as more important~\cite{sindhu2022text}; neither option would help to identify the main characteristics of simulation scenarios or the key changes that they produce.

\begin{figure}[htb]
	\centering
	\includegraphics[width=\linewidth]{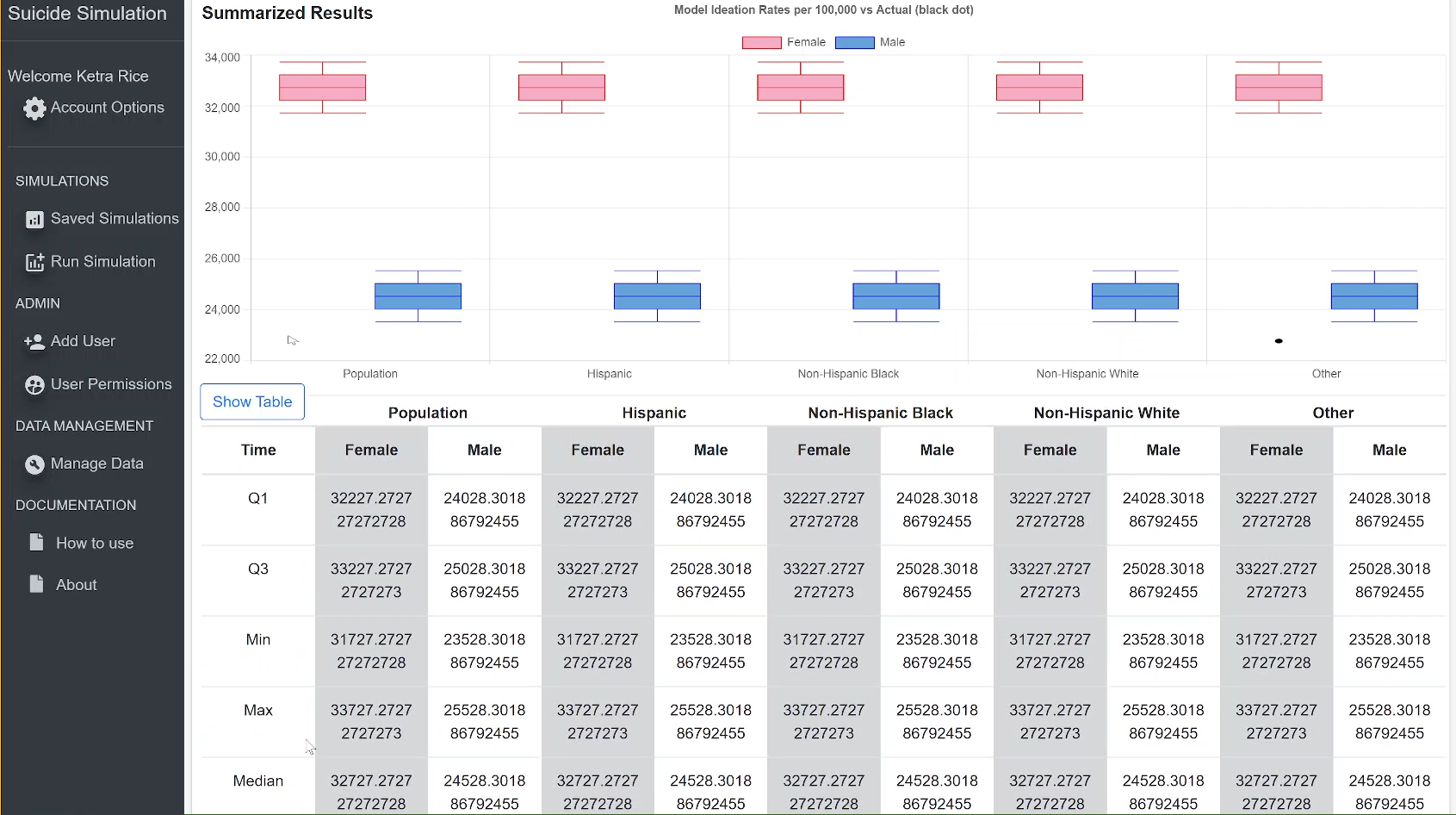}
  \caption{An Agent-Based Model platform for a U.S. federal client was presented with an emphasis on accessibility requirements \protect\shortcite{huddleston2022design}. Providing a table for each visualization contributes to meeting these requirements, as the HTML code can be used by screen readers.}
	\label{fig:results}
\end{figure}



\section{EMERGING CAPABILITIES: SIMULATION VISUALIZATION AS TEXT}
\label{sec:visualizations}

\subsection{Rationale for Using Natural Language Generation}

Since about 70\% of all human sensory receptors are in the eyes, it is no surprise that scientific visualizations are commonplace to derive insight from simulation results. New packages are regularly developed to support visualizations of simulation outputs in diverse application areas such as the large simulation datasets used in Earth system science~\shortcite{li2019vapor,wang2019flood} or molecular dynamics simulation~\cite{hildebrand2019bringing}. However, such visualizations create obstacles for individuals with visual impairments, which applies to 253 million individuals worldwide~\cite{blind}. An academic group developing a simulation software for its own purpose or research can decide to rely extensively on visualizations to interpret results. However, when a simulation software is developed for government agencies, laws on information technology can require that the simulation be accessible for people with disabilities (including visual impairments). This applies even if the software resides solely on the intranet and is intended to be used by a team that does not currently have members with visual impairments. Section 508 of the Rehabilitation Act in the United States enacts such requirements, which are echoed in the Barrier-Free Information Technology Ordinance from Germany, and various other disability discrimination acts. An approach to ensure compliance with regulations is to provide a data table for every simulation visualization (Figure~\ref{fig:results}), as screen readers can read tables one cell at a time. This \textit{technically} supports accessibility, to the same extent as a wheelchair ramp circling around a building at a steep angle would \textit{technically} provide access. Reading every simulation data point is not only cumbersome, but it also prevents users from identifying patterns, just as it would be challenging to make sense of a picture if it was read one pixel at a time. Turning visualizations into written reports focusing on the main patterns would thus broaden accessibility to simulations and ensure compliance with legal requirements.

\subsection{Core Methodological Components}
Neural networks have been used in browser extensions to automatically decode charts~\shortcite{choi2019visualizing} and newer LLMs (e.g., GPT-4) now include the ability of transforming images to text. However, this is still only an \textit{emerging} capability, hence we note that significant research efforts are still needed. Recent studies can guide modelers who seek to operate LLMs in the near future to generate textual summaries of their visualizations. First, evaluations with visually impaired individuals concluded that the preferred natural language descriptions for visualizations were \textit{very reader-specific}~\cite{lundgard2021accessible}. Modelers thus need to be cognizant of their target audience, ideally by eliciting individual preferences. Second, the usefulness of the description \textit{depends on the task} that the reader seeks to accomplish. For instance, the heatmap in Figure~\ref{fig:viz} can be examined to find the best intervention (top-right corner) or to know how the outcome depends on the two control parameters (strictly decreases as a function of both parameters). Third, users may expect different \textit{verbosity levels}. As exemplified by~\shortciteN{zong2022rich}, ''at higher verbosity the screen reader announces more structural, wayfinding content (e.g. the start and end of regions).'' More verbose summaries are not necessarily more effective, and a few bullet point statements can be a better starting point~\cite{brath2021automated}. Modelers may thus consider starting their NLG prompts by requiring a few short statements instead of a comprehensive summary. 

\section{THE NEXT FRONTIER? EXPLAINING AND FIXING SIMULATION ERRORS}
\label{sec:errors}

\subsection{Rationale for Using Natural Language Generation}

Once a simulation model can perform experiments, we need to ensure the correctness of the implementation with regards to the specification (verification) and with respect to the expected approximation of a real-world phenomenon (validation). At first, this may resemble the task of detecting and explaining implementation errors, which is increasingly studied in the nascent literature on LLMs for debugging. However, the literature on automated debugging has mostly examined the behavior of \textit{small functions}, for example by using prompts to state that a function `obviously has a bug' because it returned $Output_{prompt}$ instead of $Output_{expected}$ for a given $Input$~\shortcite{liventsev2023fully}. While locating and fixing bugs within functions is undoubtedly beneficial, verification and validation are concerned with errors that may happen at the level of the \textit{whole model}. Two sub-tasks are involved: explaining why there are errors by comparing simulation outputs with expectations, and providing guidance to address these errors. We posit that LLMs may be best employed to explain errors (e.g., `without vaccines and social distancing your population saw a reduction in COVID-19 cases but we believe that either of these interventions would have been needed to yield such results') than to identify them, which may be achieved through established statistical techniques. We also note that the guidance offered would primarily consist of generating hypotheses~\shortcite{kang2023explainable} such as `your agents may need an exposed stage before infection'. This would already be tremendously supportive for \textit{modelers in training}, by automating part of the feedback that is otherwise provided by instructors. We caution against expectations that a LLM may directly write the code logic for a model, as current LLMs write text that looks like code but does not always run and they are best suited to automate mundane tasks by writing functions that have already been encountered in their training data~\shortcite{merow2023ai}. 



\subsection{Core Methodological Components}

We view the use of LLMs to explain and address simulation errors as the next frontier, as the tools are further from practical use than in the previous sections. Two components can play an important role in allowing an LLM to identify errors. First, the LLM needs to relate the behavior of this specific model to its general knowledge base, which can leverage an LLM's ability as a causal learner. Modelers would thus need to write a prompt that states the goal of the model and summarizes its causal pathways. For example: ``A model for COVID-19 assumes that most people have not been exposed to the virus. There is a chance of getting sick upon exposure. Infected people either recover or die''. Second, the LLM would benefit from contextual information on the error (e.g., which outputs are lower/higher than expected and by how much), which can be provided by existing statistical packages that compare model outputs with expected outputs. The result can complete the prompt as follows: ``After one year, without vaccines or social distancing, nobody is infected by COVID-19 anymore. This is obviously wrong. Why?'' Note that keywords such as `obvious' can trigger LLMs such as GPT to give particular considerations to some statements~\shortcite{liventsev2023fully}. Providing guidance on the existence of the error is an arduous task. A LLM does not understand the structure of a simulation model, and it may also struggle to also understand how the code is related to the model's output since that can be an emerging behavior. It is likely that a LLM would need to be taught (by prompts) about the behavior of related models and then rely on transfer learning to investigate abnormalities in the proposed model.

\section{CONCLUSION}
\label{sec:discussion}

We focused on tasks that are enabled by the emerging technology of LLMs. There are cases in which tasks that used to be performed by other technology (e.g., question-answering systems based on information retrieval rather than machine learning) are now also using LLMs, in the manner of an \textit{oracle}. For instance, Q\&A systems have previously served to check whether the concepts and relationships of a conceptual model built by a modeling team are supported by the literature~\cite{sandhu2019social}. The same `yes' or `no' questions could be asked to a GPT-based model (e.g., `given the following documents [...], can infection from COVID-19 follow exposure to COVID-19 particles?'), but this would be a relatively minor update of a technology rather than a breakthrough for M\&S. Recently, researchers have shifted from \textit{checking} a proposed conceptual model to \textit{building} it automatically. This was first done by retrieving a corpus and repeatedly running a Q\&A system in the same manner as a facilitator would develop a model by talking with a subject matter expert~\cite{davis2022automatically}, but researchers are now examining the \textit{feasibility} of relying entirely on LLMs to build causal graphs~\shortcite{long2023can,zhang2023understanding}. We believe that this automation is an exciting step for M\&S, particularly if the conceptual model built by LLMs can then be mapped onto simulation building blocks to automatically create a working prototype~\shortcite{schroeder2022towards}. The guidance provided in this paper can thus be updated as progress is realized by the NLG community, to ensure that it ultimately benefits modeling and simulation.

\section*{ACKNOWLEDGEMENTS}
Several of the reflections in this paper are the product of fruitful discussions with numerous individuals. In particular, the author wishes to thank Mr. Anish Shrestha and Mr. Tyler Gandee, whose theses help to realize the potential but also the technical challenges in using GPT-based models. The author has also benefited from stimulating exchanges of ideas with various participants (including Dr. Ameeta Agrawal and Dr. Jose Padilla) at a research seminar co-organized at Miami University with Dr Vijay Mago. 

\footnotesize

\bibliographystyle{wsc}

\bibliography{demobib}



\end{document}